\documentclass[10pt,conference]{IEEEtran}
\usepackage{multirow}
\usepackage{mathtools}
\usepackage{graphicx}
\usepackage{amsmath}
\usepackage{color}
\usepackage{epsfig}
\usepackage{amsfonts}
\usepackage{amssymb}
\usepackage{amsthm}
\usepackage[usenames,dvipsnames]{pstricks}
\usepackage{epstopdf}
\usepackage{algorithm}
\usepackage{caption}
\usepackage{subcaption}
\usepackage[noend]{algpseudocode}
\newcommand{\argmax}{\mathop{\text{argmax}}}

\begin{document}
\title{{\Huge Multiple Access in the Delay-Doppler Domain using OTFS 
modulation}}
\author{G. D. Surabhi, Rose Mary Augustine, and A. Chockalingam \\
Department of ECE, Indian Institute of Science, Bangalore 560012 
\vspace{-5mm}}
\maketitle
\begin{abstract}
Orthogonal time frequency space (OTFS) modulation is a recent modulation 
scheme designed in the delay-Doppler domain. It has been shown to achieve 
superior performance compared to conventional multicarrier modulation 
schemes designed in the time-frequency domain. In this paper, we consider 
OTFS based multiple access (OTFS-MA), where delay-Doppler bins serve as 
the resource blocks for multiple access. Different delay-Doppler resource 
blocks (DDRBs) in the delay-Doppler grid are allocated to different users 
for multiple access. We consider three different DDRB allocation schemes. 
While Scheme 1 multiplexes the users along the delay axis, Scheme 2 
multiplexes them along the Doppler axis. In both these schemes, each 
user's signal spans the entire time-frequency plane. Scheme 3 allocates 
the DDRBs in such a way that each user's signal is limited to span only 
over a subset of the time-frequency plane. We study the performance of 
OTFS-MA in high mobility environments on the uplink and compare it with 
those of OFDMA and SC-FDMA. Our results show that OTFS-MA (with 
maximum-likelihood detection in small dimension systems and with a 
message passing based detection in large dimension systems) achieves 
better performance compared to OFDMA and SC-FDMA. We also present the 
performance of a multiuser channel estimation scheme using pilot symbols 
placed in the delay-Doppler grid.
\end{abstract}

\section{Introduction}
\label{sec1}
\let\thefootnote\relax\footnote{This work was supported in part by the
J. C. Bose National Fellowship, Department of Science and Technology,
Government of India, and the Intel India Faculty Excellence Program.}
Next generation wireless systems are envisioned to support high speed 
communications with energy efficiency and high reliability in various 
wireless environments. Enabling high speed and reliable communication
in high mobility scenarios, which arise in environments such as 
high-speed trains, vehicle-to-vehicle, and vehicle-to-infrastructure 
communications, requires techniques which are specially suited for 
the dynamic nature of wireless channels. The wireless channels in such 
scenarios are rapidly time varying and hence doubly dispersive in nature, 
with the multipath effects causing time dispersion and Doppler shifts 
causing frequency dispersion \cite{jakes}. Conventional multicarrier 
modulation techniques are primarily designed to combat the multipath 
effects that cause inter-symbol interference (ISI) \cite{Proakis}. 
However, high mobility or the use of high frequency carriers (e.g., 
mmWave frequencies) in low to medium mobility environments results in 
Doppler shift causing inter-carrier interference (ICI), which degrades 
the performance of conventional multicarrier modulation schemes.

Orthogonal time frequency space modulation (OTFS) is a new modulation
technique suited for doubly dispersive wireless channels. OTFS was first 
introduced in \cite{otfs1}, where it was shown to outperform conventional 
multicarrier modulation schemes such as OFDM in channels with high Doppler 
spreads. The robustness of OTFS modulation in high mobility environments 
(e.g., vehicle speed as high as 500 km/h) and mmWave communication 
environments has been demonstrated in \cite{otfs1}-\cite{mmwave_otfs}. 
The basic idea behind OTFS modulation can be briefly explained as follows. 
OTFS is a 2-dimensional (2D) modulation technique which uses the 
delay-Doppler domain for multiplexing information symbols. This is in 
contrast to conventional multicarrier modulation schemes which multiplex 
symbols in the time-frequency domain. OTFS modulation uses a series of 
2D transformations by which the rapidly time varying channel is converted  
into a slowly varying channel in the delay-Doppler domain. The slow 
variability of the delay-Doppler channels reduces the overhead of 
frequent channel estimation in channels with small coherence time. 
Also, these transformations are such that all the information symbols 
are coupled to the channel in the delay-Doppler domain in the same 
fashion. This greatly simplifies the equalizer design in rapidly time 
varying channels. Another attractive feature of OTFS is that it could 
be architected with pre- and post processing operations over any 
existing multicarrier system. 

Recognizing the superior performance and implementation simplicity of 
OTFS, several works  studying various aspects of OTFS have emerged recently
\cite{ofdm_otfs1}-\cite{emb_pil}. A linear vector channel model for 
OTFS has been derived and a low-complexity message passing based OTFS 
signal detection scheme has been proposed in \cite{otfs4}. Another 
low-complexity OTFS signal detection scheme based on Markov chain Monte 
Carlo technique has been proposed in \cite{otfs5}. Low-complexity 
implementation of OTFS over conventional OFDM systems has been reported 
in \cite{ofdm_otfs1},\cite{ofdm_otfs2}. OTFS modulation in MIMO 
communication settings (MIMO-OTFS) with a focus on MIMO-OTFS signal 
detection and channel estimation has been reported in \cite{mimo_otfs}. 
A diversity order analysis for OTFS has been presented in \cite{div_otfs}, 
where it has been shown that the asymptotic diversity order of OTFS (as 
SNR $\rightarrow \infty$) is one, and that, in the finite SNR regime, 
potential for a higher diversity slope is witnessed before the diversity 
one regime takes over. Space-time coding to achieve full spatial and 
delay-Doppler diversity in MIMO-OTFS systems is proposed in \cite{stc_otfs}. 
In \cite{puls_otfs}, the performance of OTFS with practical pulse shaping
has been considered. A framework that relates the generalized frequency
division multiplexing (GFDM) and OTFS has been formulated in 
\cite{gfdm_otfs} and a bit error performance comparison showed that OTFS 
performs better than GFDM. 

In this paper, we consider OTFS modulation for multiuser communication 
on the uplink, where users are multiplexed on the delay-Doppler grid
which is designed by considering the maximum delay and Doppler spreads 
of the multiuser channel. In this multiple access system, called as
OTFS-MA (OTFS multiple access), bins in the delay-Doppler grid serve 
as the resource blocks. These resource blocks are called the
DDRBs (delay-Doppler resource blocks). Different DDRBs are allocated 
to different users for multiple access. We consider three 
different DDRB allocation schemes. While Scheme 1 multiplexes the users 
along the delay axis, Scheme 2 multiplexes them along the Doppler axis. 
In both these schemes, each user's signal spans the entire time-frequency 
plane. Scheme 3 allocates the DDRBs in such a way that each user's signal 
is limited to span over only a subset of the time-frequency plane. All 
these three schemes have been suggested in \cite{otfs2}. The sum rate of
Scheme 3 has been analyzed in \cite{ma_otfs}. Here, we study the bit
error performance of OTFS-MA with the above allocation schemes in high 
mobility environments on the uplink 
and compare it with those of other popular multiple access schemes such as
OFDMA and SC-FDMA. Our results show that OTFS-MA (with maximum-likelihood 
detection in small dimension systems and with a message passing based 
detection in large dimension systems) achieves better performance compared 
to OFDMA and SC-FDMA. We also present the performance of a multiuser channel 
estimation scheme using pilot symbols placed in the delay-Doppler grid.

The rest of this paper is organized as follows. In Sec. \ref{sec2}, the
OTFS-MA system model and the DDRB allocation schemes considered are 
presented. In Sec. \ref{sec3}, the performance of OTFS-MA under the
considered allocation schemes are compared. A comparison between 
OTFS-MA, OFDMA, and SC-FDMA with ML detection is also presented.
In Sec. \ref{sec4}, a message passing based detection and its 
performance are presented. A channel estimation technique for OTFS-MA 
and its performance are presented in Sec. \ref{sec5}. Conclusions are 
presented in Sec. \ref{sec6}.

\begin{figure}[t]
\centering
\includegraphics[width=7.5 cm, height=8.0 cm]{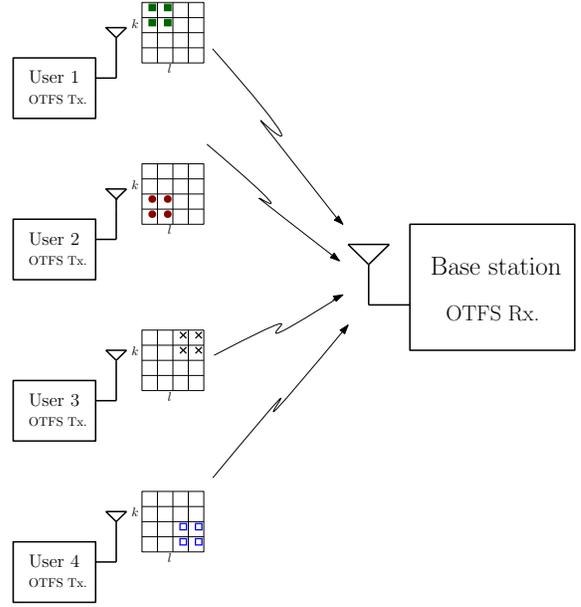}
\vspace{2mm}
\caption{OTFS multiple access (OTFS-MA) on the uplink.}
\label{OTFS_UL}
\vspace{-2mm}
\end{figure}

\section{OTFS-MA system model}
\label{sec2}
\subsection{Uplink OTFS-MA system model}
\label{sec2A}
Consider an OTFS-MA  system  with $K_u$ uplink users communicating with 
a base station (BS) as shown in Fig. \ref{OTFS_UL}. Each user employs 
OTFS modulation for signaling on the uplink. Each user is equipped with 
a single antenna transmitter and the BS is equipped with a single antenna 
receiver. In OTFS, information symbols are multiplexed in the delay-Doppler 
domain, i.e., the information symbols are multiplexed on an $N\times M$ 
delay-Doppler grid which is denoted by $\Gamma$, and is given by
\begin{equation}
\Gamma=\lbrace (\tfrac{k}{NT},\tfrac{l}{M\Delta f}), k=0,1, \cdots,
N-1, l=0,1, \cdots, M-1\rbrace.
\label{DDgrid}
\end{equation}
Here, $1/NT$ and $1/M\Delta f$ represent the quantization steps
of the Doppler shift and the delay, respectively, so that $N$ and $M$ 
denote the number of Doppler and delay bins, respectively. Let
$\tau_{\mbox{\scriptsize{max}}}$ and $\nu_{\mbox{\scriptsize{max}}}$ 
denote the maximum delay and Doppler spread of the multiuser channel, 
respectively. 
Then, $\Delta f$ must be such that $\nu_{\mbox{\scriptsize{max}}}
<\Delta f < 1/\tau_{\mbox{\scriptsize{max}}}$. We refer to a bin on the 
delay-Doppler grid $\Gamma$ in \eqref{DDgrid} as a delay-Doppler resource 
block (DDRB). Let $x_u[k,l]$, $k=0,1,\cdots N-1$, $l=0,1,\cdots M-1$, and 
$u=0,1,\cdots K_u-1$ denote the information symbol from a modulation 
alphabet $\mathbb{A}$ (e.g., QAM/PSK) transmitted by the $u$th user on 
the $(k,l)$th DDRB. 

The information symbols of the $u$th user, i.e., $x_u[k,l]$s, in the 
delay-Doppler domain are mapped to the TF domain using the inverse 
symplectic finite Fourier transform (ISFFT) and windowing. Assuming 
rectangular windowing, the modulated TF signal corresponding to the 
$u$th user is given by
\begin{equation}
X_u[n,m] = \frac{1}{\sqrt{MN}}\sum_{k=0}^{N-1}\sum_{l=0}^{M-1} x_u[k,l]e^{j2\pi
\left(\frac{nk}{N}-\frac{ml}{M}\right)}.
\label{otfsmod}
\end{equation}
The TF signal so obtained is converted into a time domain signal for 
transmission using Heisenberg transform with a transmit pulse denoted 
by $g_{tx}(t)$. The transmitted time domain signal of the $u$th user 
therefore is given by
\begin{equation}
x_u(t)= \sum_{n=0}^{N-1} \sum_{m=0}^{M-1} X_u[n,m]g_{tx}(t-nT)e^{j2\pi m 
\Delta f (t-nT)}.
\label{tfmod}
\end{equation}
The transmitted signal $x_u(t)$ passes through the channel whose complex 
baseband channel response in the delay-Doppler domain is denoted by 
$h_u(\tau,\nu)$. The received time domain signal $y(t)$ at the BS is 
given by
\begin{equation}
\label{channel}
y(t)=\sum_{u=0}^{K_u-1}\int_{\nu} \int_{\tau} h_u(\tau,\nu)x(t-\tau)e^{j2\pi\nu(t-\tau)} \mathrm{d} \tau \mathrm{d} \nu+v(t),
\end{equation} 
where $h_u(\tau,\nu)$ is the delay-Doppler channel between the $u$th
user and the BS and $v(t)$ denotes the additive white Gaussian noise 
at the BS receiver. The received signal at the BS is matched filtered 
with a receive pulse $g_{rx}(t)$, yielding the cross-ambiguity 
function denoted by $A_{g_{rx},y}(t,f)$ and given by
\begin{equation}
\label{crossambig}
A_{g_{rx},y}(t,f)=\int g_{rx} ^*(t'-t) y(t') e^{-j2 \pi f(t'-t)} \mathrm{d}t'.
\end{equation}
The pulses $g_{tx}(t)$ and $g_{rx}(t)$ are chosen such that the 
biorthogonality condition is satisfied, i.e., 
$A_{g_{rx},g_{tx}}(t,f)|_{nT, m\Delta f}=\delta (m)\delta (n)$. 
Sampling $A_{g_{rx},y}(t,f)$ at $t =nT$ and $f=m \Delta f$ yields 
the matched filter output, given by
\begin{equation}
\label{wigner}
Y[n,m] = A_{g_{rx},y}(t,f)|_{t=nT,f =m \Delta f}.
\end{equation} 
Finally, $Y[n,m]$  is converted from TF domain back to delay-Doppler 
domain to obtain $y[k,l]$ as
\begin{equation}
y[k,l]=\frac{1}{\sqrt{MN}}\sum_{k=0}^{N-1}\sum_{l=0}^{M-1}
Y[n,m]e^{-j2\pi\left(\frac{nk}{N}-\frac{ml}{M}\right)}.
\label{otfsdemod}
\end{equation}
If $h_u(\tau,\nu)$ has finite support bounded by
$(\tau_{\mbox{\scriptsize{max}}},\nu_{\mbox{\scriptsize{max}}})$ and if 
$A_{g_{rx}g_{tx}}(t,f)=0$ for $t \in (nT-\tau_{\mbox{\scriptsize{max}}},nT+\tau_{\mbox{\scriptsize{max}}})$, 
$f \in (m\Delta f-\nu_{\mbox{\scriptsize{max}}},m\Delta f+\nu_{\mbox{\scriptsize{max}}})$, 
$\forall (n,m) \neq (0,0)$, the end-to-end input-output relation for the 
considered uplink OTFS-MA system can be written as 
\begin{align}
y[k',l']=&\frac{1}{MN}\sum_{u=0}^{K_u-1}\sum_{k=0}^{N-1}
\sum_{l=0}^{M-1}x_u[k,l] \nonumber \\
& .\tilde{h}_u[(k'-k)_N,(l'-l)_M]+v[k',l'],
\label{2Dconv}
\end{align}
where $(.)_N$ denotes modulo-$N$ operation, $v[k,l]$ denotes the 
additive white Gaussian noise, and $\tilde{h}_u(k,l)$ is the sampled 
version of the impulse response function $\tilde{h}_u(\nu, \tau)$,
which is the circular convolution of $h_u(\tau,\nu) $ with the window 
function in the delay-Doppler domain, at $\nu=\frac{k}{NT}$ and 
$\tau=\frac{l}{M\Delta f}$ \cite{otfs1}.

Consider that the channel between the $u$th user and the BS, i.e., 
$h_u(\tau,\nu)$, has $P_u$ paths, where $h_{u,i}$, $\tau_{u,i}$, 
$\nu_{u,i}$ denote the channel gain, delay, and Doppler shift, 
respectively, associated with the $i$th path of the $u$th user. The 
$u$th user's channel in the delay-Doppler domain is then given by
\begin{equation}
h_u(\tau,\nu) =\sum_{i=1}^{P_u} h_{u,i} \delta(\tau -\tau_{u,i}) \delta(\nu-\nu_{u,i}),
\label{sparsechannel}
\end{equation}
where $h_{u,i}$s are assumed to be i.i.d.  Let $\tau_{u,i}\triangleq\frac{\alpha_{u,i}}{M\Delta f}$ 
and $\nu_{u,i}\triangleq\frac{\beta_{u,i}+b_{u,i}}{NT}$, where $\alpha_{u,i}$, 
$\beta_{u,i}$ are integers and $-\frac{1}{2}<b_{u,i} \leq \frac{1}{2}$ is the 
fractional Doppler corresponding to $\nu_{u,i}$. Fractional delays are not
considered since the sampling time (delay resolution $1/M\Delta f$)
is typically small in wideband systems and hence it can be approximated
to the nearest sampling point \cite{DTse}. With this, the input-output 
relation is given by
\begin{align}
\small
y[k,l] &= \sum_{u=0}^{K_u-1}\sum_{i=1}^{P_u}\sum_{q'=0}^{N-1}
\left(\frac{e^{-j2\pi(-q'-b_{u,i})}-1}
{Ne^{-j\frac{2\pi}{N}(-q'-b_{u,i})}-N}\right) h_{u,i} \nonumber \\
& . e^{-j2\pi \tau_{u,i}
\nu_{u,i}}x_u[(k-\beta_{u,i}+q')_N,(l-\alpha_{u,i})_M]+v[k,l].
\label{inpoutfracdop}
\end{align}

The 2D circular convolution of symbols transmitted by each user with
the corresponding channel in \eqref{2Dconv} can be written in a 
vectorized form as in the case of single user setting \cite{otfs4}. 
Denoting the OTFS symbol
vector transmitted by $u$th user by $\mathbf{x}_u \in \mathbb{C}^{MN\times 1}$
($\mathbf{x}_{u_{\;k+Nl}}=x_u[k,l]$) and the channel matrix of $u$th user by 
$\mathbf{H}_u \;\in \mathbb{C}^{MN\times MN}$, the input-output relation 
in multiuser OTFS can be written as
\begin{align}
\mathbf{y}=&\sum_{u=0}^{K_u-1}\mathbf{H}_u\mathbf{x}_u+\mathbf{v}, \nonumber \\
=&[\mathbf{H}_1 \mathbf{H}_2 \cdots \mathbf{H}_{K_u}]\begin{bmatrix}
\mathbf{x}_{1} \\
\mathbf{x}_{2}\\
\vdots\\
\mathbf{x}_{K_u}
\end{bmatrix}+\mathbf{v}, 
\label{vecform}
\end{align}
where $\mathbf{y}\in \mathbb{C}^{MN\times 1}$ is the received vector at 
the BS, and $\mathbf{v}$ is the additive white Gaussian noise vector with 
$\mathbf{v}_{k+Nl}=v[k,l]$.

\subsection{DDRB allocation schemes}
\label{sec2a}
In this subsection, we present three different schemes for allocation 
of DDRBs to users in an uplink OTFS-MA system.
\subsubsection{Scheme 1 (Multiplexing users along the delay axis)} 
\label{sec2a1}
\begin{figure}
\centering
\includegraphics[width=5 cm, height=5.0 cm]{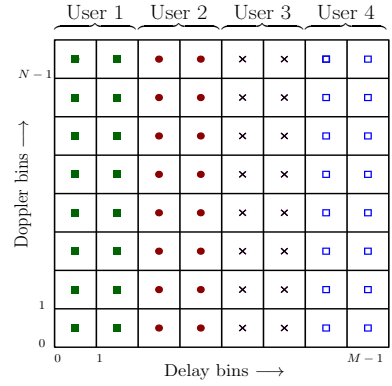}
\vspace{-0mm}
\caption{DDRB allocation in an $N\times M$ delay-Doppler grid in Scheme 1.}
\label{muxoverM}
\vspace{-4mm}
\end{figure}

In this scheme, disjoint and contiguous bins along the delay axis are 
allocated to each user such that each user gets $M/K_u$ columns of the 
delay-Doppler grid for transmission (see Fig. \ref{muxoverM}). The 
delay-Doppler grid of the $u$th user will have
\begin{equation}
x_u[k,l]=\begin{cases}
a\in \mathbb{A}  & \textnormal{if} \;\; k\in \lbrace 0,1\cdots N-1\rbrace \;
\&  \\
& l\in \lbrace u\frac{M}{K_u},\cdots (u+1) \frac{M}{K_u}-1 \rbrace \\
0  & \textnormal{otherwise} .
\end{cases}
\end{equation}
Figure \ref{muxoverM} shows an example of Scheme 1 allocation, where
an $N\times M=8\times 8$ delay-Doppler grid gets allocated to four users. 
Note that, although the users transmit on non-overlapping DDRBs, the 
symbols transmitted by each user experience multiuser interference 
(MUI) due to the 2D circular convolution operation in \eqref{2Dconv}.
The amount of MUI experienced depends on the delay spread of the channels.
Hence, the received signal at the BS has to be jointly decoded. A way
to receive MUI free signal at the BS using Scheme 1 is to use a set of 
DDRBs as guard bands in the delay domain, based on the delay spread of the 
adjacent users' channels \cite{ma_patent}. However, this reduces the 
spectral efficiency of the overall system, especially in the channels 
with large delay spreads which require large guard bands for MUI-free 
reception. 
 
\subsubsection{Scheme 2 (Multiplexing users along the Doppler axis)}
\label{sec2a2}
\begin{figure}
\centering
\includegraphics[width=5cm, height=4.5cm]{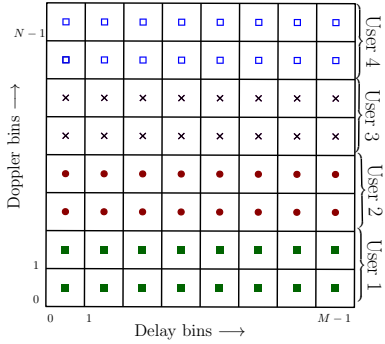}
\vspace{-0mm}
\caption{DDRB allocation in an $N\times M$ delay-Doppler grid in Scheme 2.}
\label{muxoverN}
\vspace{-2mm}
\end{figure}

In this scheme, non-overlapping and contiguous DDRBs along the Doppler
axis are allocated to each user such that each user gets $N/K_u$
rows of the delay-Doppler grid for transmission (see Fig. \ref{muxoverN}). 
The delay-Doppler grid of the $u$th user will have
\begin{equation}
x_u[k,l]=\begin{cases}
a\in \mathbb{A} & \textnormal{if} \;\; k\in \lbrace u\frac{N}{K_u},\cdots (u+1) 
\frac{N}{K_u}-1\rbrace \; \& \\
& l\in \lbrace 0,1\cdots M-1\rbrace \\
0  & \textnormal{otherwise} .
\end{cases}
\end{equation}
Figure \ref{muxoverN} shows an example of Scheme 2 allocation, where an
$N\times M=8\times 8$ delay-Doppler grid gets allocated to four users.
In Scheme 2 also, the 2D circular convolution operation in \eqref{2Dconv} 
results in the symbols transmitted by each user to experience MUI, 
requiring the BS to jointly decode the symbols corresponding to all 
the users. Allowing guard bands along the Doppler domain can result in 
MUI-free reception at the BS \cite{ma_patent}. However, for channels 
with high Doppler spread, this may result in reduced spectral efficiency 
of the system.
\begin{figure}
\centering
\includegraphics[width=6cm, height=5cm]{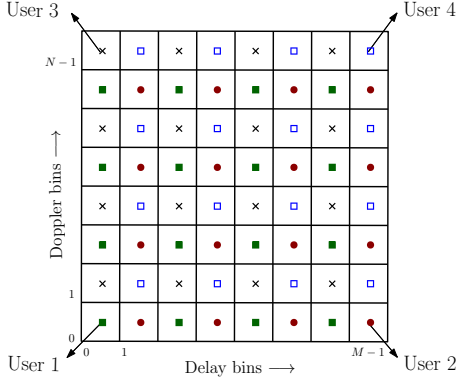}
\vspace{-0mm}
\caption{DDRB allocation in an $N\times M$ delay-Doppler grid in Scheme 3 
\cite{ma_otfs}.}
\label{IF}
\vspace{-4mm}
\end{figure}

\subsubsection{Scheme 3 (Allocation scheme in \cite{ma_otfs})}
\label{sec2a3}
In Schemes 1 and 2, each user's signal spans the entire time-frequency 
plane. In Scheme 3 \cite{ma_otfs}, the allocation of DDRBs is done in 
such a way that each user's signal can be restricted to span only over 
a subset of the time-frequency plane. 
The allocation is such that $MN/K_u$ symbols 
corresponding to a given user are placed at equal intervals in the delay 
as well as Doppler domains (see Fig. \ref{IF}). These intervals are 
determined by two parameters denoted by $g_1$ and $g_2$ such that 
$K_u=g_1g_2$, with $M=\kappa_1 g_1$ and $N=\kappa_2 g_2$, where 
$\kappa_1, \kappa_2 \in \mathbb{Z}_+$. The allocation is such that the 
delay-Doppler grid corresponding to the $u$th user will have
\begin{equation}
x_u[k,l]=\begin{cases}
a\in \mathbb{A}  & \textnormal{if} \; k=\lfloor u/g_1\rfloor+g_2p \; \& \\
& l=(u)_{g_1}+g_1q  \\
0  & \textnormal{otherwise} .
\end{cases}
\end{equation} where $p \in\lbrace 0,1,\cdots N/g_2-1 \rbrace$ and $q \in
\lbrace 0,1,\cdots,M/g_1-1\rbrace$. 
This scheme results in a periodic interleaving of symbols from each user as 
shown in Fig. \ref{IF} for a system with $M=N=8$, $K_u=4$, and $g_1=g_2=2$. 
It has been shown in \cite{ma_otfs} that, with this allocation, the 
time-frequency symbols $X_u[n,m]$ corresponding to $u$th user can be 
restricted to a region $[(NT/g_2)(u)_{g_2},(NT/g_2)((u)_{g_2}+1)]$ in time 
and $[(M/g_1)\lfloor u/g_2\rfloor \Delta f,(M/g_1)(\lfloor u/g_2\rfloor+1) 
\Delta f]$ in frequency. These regions are non-overlapping in the TF plane, 
and hence it enables the BS to separate out the received TF signal of each 
user. At the BS, the TF signal of $u$th user, denoted by $Y_u[n,m]$, is 
transformed back to the delay-Doppler domain through SFFT as \cite{ma_otfs}
\begin{equation}
\small
y_u[k',l']=\frac{1}{\sqrt{MN}}\sum_{n=0}^{N/g_2-1}\sum_{m=0}^{M/g_1-1}
Y_u[n,m]e^{-j2\pi\left(\frac{nk'}{N/g_2}-\frac{ml'}{M/g_1}\right)}.
\label{IF_SFFT}
\end{equation}
The SFFT in \eqref{IF_SFFT} results in the $u$th user's signal in 
delay-Doppler domain $y_u[k',l']$, $k'=0,1,\cdots,N/g_2-1$, 
$l'=0,1,\cdots,M/g_1-1$. 
Note that the SFFT computation in \eqref{IF_SFFT} is over the region in 
TF domain to which the $u$th user's signal is restricted to. This is 
unlike the SFFT computation in \eqref{otfsdemod}, which involved 
computing SFFT over the entire TF plane. This difference in SFFT 
computation results in a slightly different input-output relation for 
Scheme 3 compared to those of Schemes 1 and 2. The input-output relation 
for Scheme 3 has been derived in \cite{ma_otfs} and is given by
\begin{align}
y_u[k',l']=&\sum_{k=0}^{N/g_2-1}\sum_{l=0}^{M/g1-1}\tilde{x}_u[k,l] \nonumber \\
& \hat{h}_u[(k'-k)_{N/g_2},(l'-l)_{M/g_1}]+v_q[k',l'],
\label{2Dconv1}
\end{align}
where $\tilde{x}_u[p,q]\triangleq x_u(k=\lfloor u/g_1\rfloor+g_2p,
l=(u)_{g_1}+g_1q)$ and $v_q[k',l']\sim \mathcal{CN}(0,1/(g_1g_2))$, and 
\begin{align}
\hat{h}_u[r,s]=&\sum_{i=1}^{P_u}\big[h_{u,i}e^{-j2\pi(\nu_{u,i}\tau_{u,i}
+\frac{\tau_{u,i}}{T}\frac{M}{g_1}\lfloor \frac{u}{g_2}\rfloor -
\frac{\nu_{u,i}}{\Delta f}\frac{N}{g_2}(u)_{g_2})} \nonumber \\
& \; \; \mathcal{F}_{u,i}[s] \mathcal{G}_{u,i}[r]\big], \nonumber \\
\mathcal{F}_{u,i}[s]=& \frac{1}{M}\sum_{m=0}^{M/g_1-1}e^{-j2\pi m
\Big(\frac{(u)_{g_1}}{M}-\frac{s}{M/g_1}+\frac{\tau_{u,i}}{T}\Big)}, \nonumber \\
\mathcal{G}_{u,i}[r]=& \frac{1}{N}\sum_{n=0}^{N/g_2-1}e^{j2\pi n
\Big(\frac{\lfloor u/g_1 \rfloor}{N}-\frac{r}{N/g_2}+\frac{\nu_{u,i}}{\Delta f}\Big)}.
\end{align}
The 2D convolution in \eqref{2Dconv1} can be vectorized as
\begin{equation}
\mathbf{y}_u=\hat{\mathbf{H}}_u\tilde{\mathbf{x}}_u+\tilde{\mathbf{v}}_u,
\label{hx_if}
\end{equation}
where $\tilde{\mathbf{x}}_u\in \mathbb{C}^{MN/K_u\times 1}$ and 
$\hat{\mathbf{H}}_u\in \mathbb{C}^{MN/K_u\times MN/K_u}$.
Since users' signals at the BS in this scheme are separable in the 
TF plane, the TF signal corresponding to each user can be individually
mapped to the delay-Doppler plane for detection. This leads to reduced 
detection complexity at the BS.

\section{ML detection performance results}
\label{sec3}
In this section, we present the bit error rate (BER) performance of 
uplink OTFS-MA under ML detection. We compare the performance of the 
DDRB allocation schemes presented in Sec. \ref{sec2a}. We also compare 
the BER performance of OTFS-MA with those of OFDMA and SC-FDMA.

\begin{figure}[t]
\centering
\includegraphics[width=9.5cm, height=6.5cm]{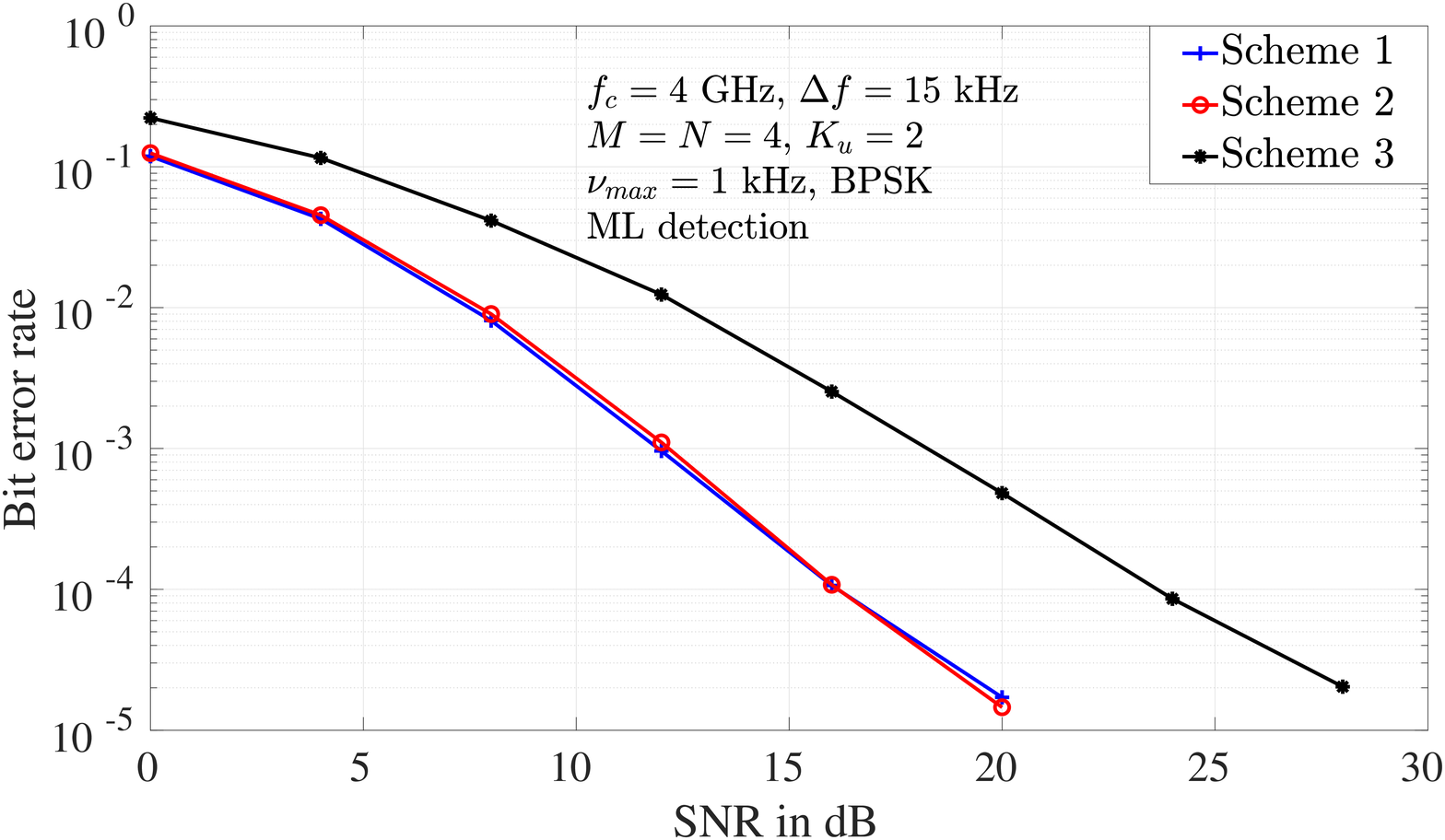}
\vspace{-3mm}
\caption{BER performance of uplink OTFS-MA with different DDRB 
allocation schemes with $M=N=4$, $K_u=2$, and ML detection.}
\vspace{-4mm}
\label{ber_ml0}
\end{figure}

{\em Performance of DDRB allocation Schemes 1,2,3:}
Figure \ref{ber_ml0} shows the BER performance of uplink OTFS-MA with 
the different DDRB allocation schemes discussed in Sec.\ref{sec2a}. A 
delay-Doppler grid with $M=N=4$ is considered. The delay-Doppler bins 
in this grid are shared among $K_u=2$ users. A carrier frequency of 4 GHz, 
subcarrier spacing of 15 kHz, and BPSK modulation are used. A four-tap 
delay-Doppler channel $(P_u=4, \forall u)$ with exponential power delay 
profile and 
Jakes Doppler spectrum \cite{jakes_spec} is considered for all the users. 
The Doppler shift corresponding to the $i$th tap of $u$th user is generated 
using $\nu_{u,i}={\nu}_{\mbox{\scriptsize{max}}}\cos(\theta_{u,i})$, where 
${\nu}_{\mbox{\scriptsize{max}}}$ is the maximum Doppler shift which is 
taken to be 1 kHz for all the users and $\theta_{u,i}$ is uniformly 
distributed over $[-\pi,\pi]$. From Fig. \ref{ber_ml0}, we observe that 
the BER performance of OTFS-MA using the allocation Scheme 1 (in Sec. 
\ref{sec2a1}) and Scheme 2 (in Sec. \ref{sec2a2}) is nearly the same and 
is superior compared to that of Scheme 3  (in Sec. \ref{sec2a3}). This 
can be explained as follows. In Scheme 3, each user's symbols are 
allowed to spread only in a restricted and disjoint region in the 
time-frequency plane, whereas the symbols in Schemes 1 and 2 are allowed 
to spread over the entire TF plane. In Scheme 3, the restricted spreading 
of each user's signal in the TF plane when brought back to the delay-Doppler 
plane through a reduced point SFFT operation hurts the bit error 
performance. Whereas, in Schemes 1 and 2, the spreading of each user's
signal over the entire TF plane and the full point SFFT operation 
to bring back this TF signal to the delay-Doppler plane followed by 
joint detection of all users' symbols result in improved performance 
compared to that of Scheme 3. An issue with the joint detection is its 
high complexity. We address this issue in Sec. \ref{sec4} where a 
low complexity joint detection scheme is proposed using message passing 
approach.

\begin{figure}
\centering
\includegraphics[width=9.5cm, height=6.5cm]{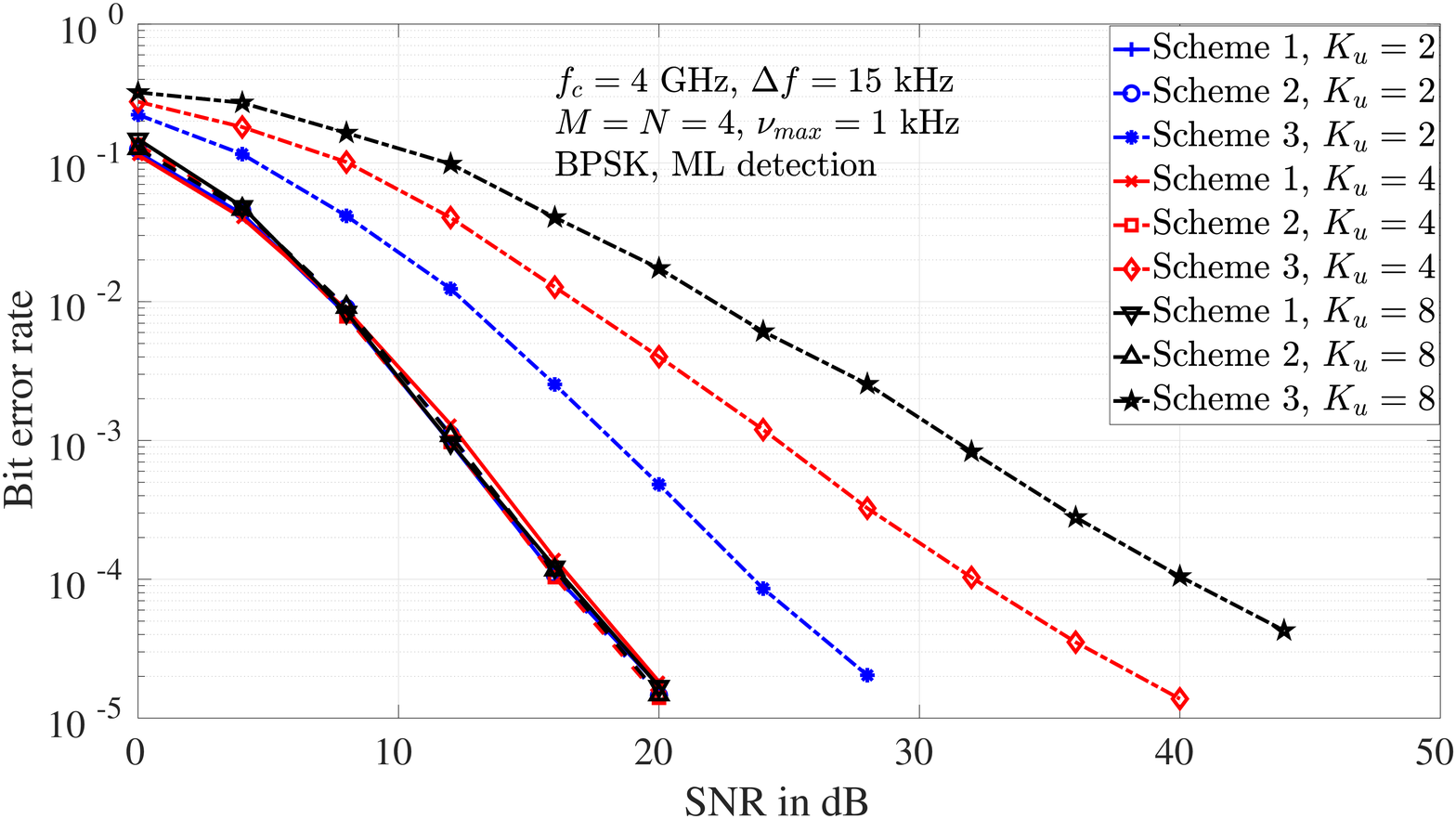}
\vspace{-3mm}
\caption{ BER performance of uplink OTFS-MA with different DDRB 
allocation schemes with $M=N=4$, $K_u=2,4,8$, and ML detection.}
\vspace{-4mm}
\label{ber_ml1}
\end{figure}

{\em Effect of number of uplink users:}
In Fig. \ref{ber_ml1}, we plot the BER performance of OTFS-MA with DDRB 
allocation Schemes 1, 2, and 3, for $K_u=2,4,8$. All the other parameters 
are the same as those used in Fig. \ref{ber_ml0}. From Fig. \ref{ber_ml1}, 
it can be seen that Schemes 1 and 2 show nearly the same performance with 
increase in the number of uplink users due to joint detection. Also, 
Schemes 1 and 2 outperform Scheme 3. It can be seen that, unlike Schemes 
1 and 2, the 
BER performance with Scheme 3 degrades with the increase in the number 
of uplink users. This can be explained as follows. As mentioned before,
the transmitted TF signal of each user in Scheme 3 is restricted to a 
specific region in the TF plane. The size of this region in the TF plane 
over which the symbols are spread is inversely proportional to the number 
of users. Therefore, increase in the number of users for a given $M$ and 
$N$ reduces the spread in the TF plane for each user, which degrades the 
performance of the system.

\begin{figure}
\centering
\includegraphics[width=9.5cm, height=6.5cm]{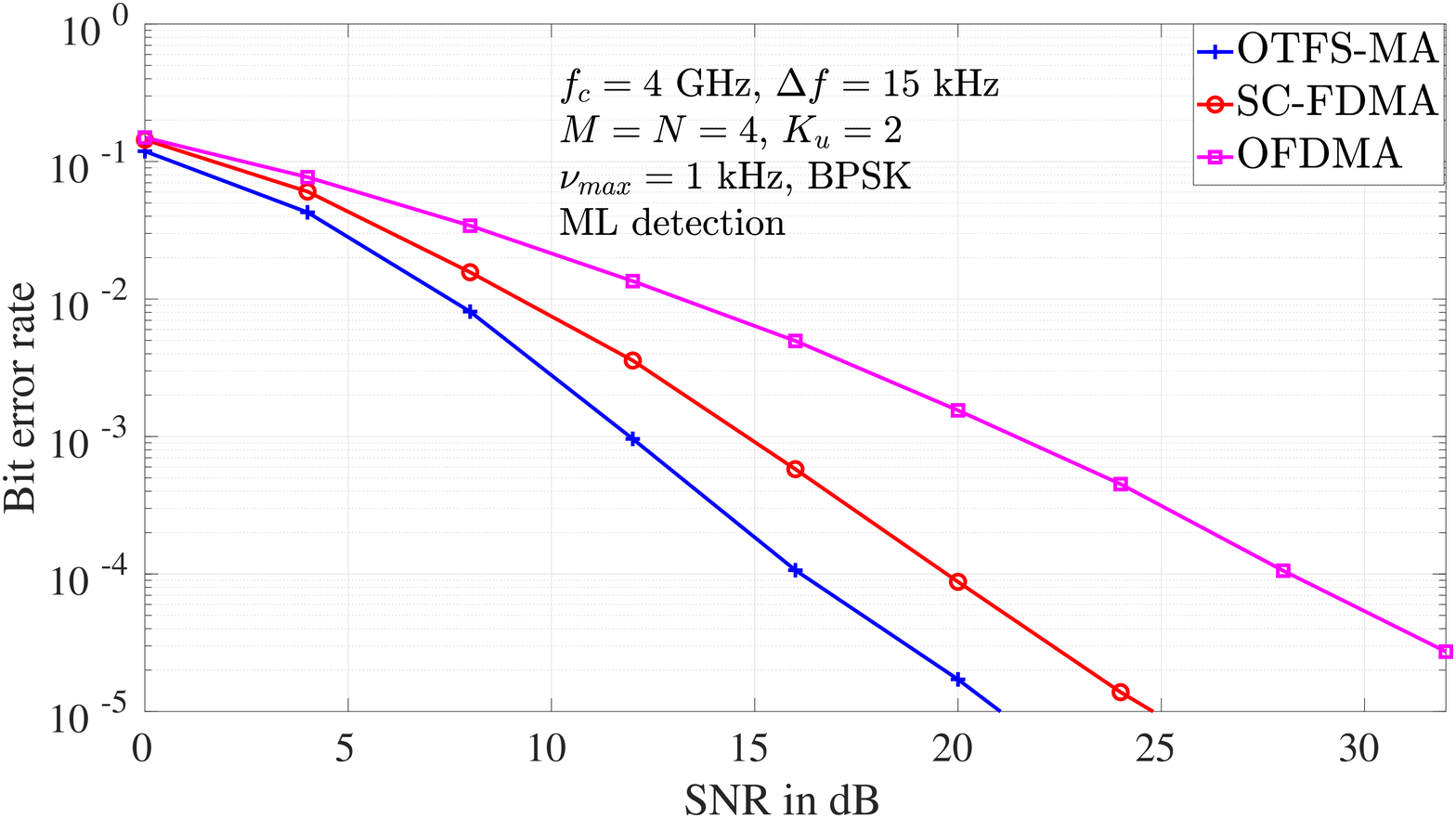}
\vspace{-3mm}
\caption{BER performance comparison between OTFS-MA, OFDMA, and SC-FDMA
with ML detection.}
\vspace{-4mm}
\label{ber_ml2}
\end{figure}

{\em Comparison between OTFS-MA, OFDMA, and SC-FDMA:}
Figure \ref{ber_ml2} shows a BER performance comparison between OTFS-MA  
with Scheme 1 allocation, OFDMA, and SC-FDMA. As before, a carrier frequency 
of 4 GHz, a subcarrier spacing of 15 kHz, exponential power delay profile, 
and Jakes Doppler spectrum are considered. For all the three systems, joint 
ML detection is used at the BS. The maximum Doppler considered is 1 kHz, 
which corresponds to a speed of 270 km/h at 4 GHz carrier frequency. The 
Doppler shift corresponding to the $i$th tap of $u$th user's channel is 
generated using $\nu_{u,i}=\nu_{\mbox{\scriptsize{max}}}\cos(\theta_{u,i})$,
where $\nu_{\mbox{\scriptsize{max}}}$ is the maximum Doppler shift and
$\theta_{u,i}$ is uniformly distributed over $[-\pi,\pi]$. From Fig.
\ref{ber_ml2}, it can be seen that the performance of OTFS-MA is superior 
compared to the performance of both OFDMA and SC-FDMA. For example, OTFS-MA 
achieves an SNR gain of about 4 dB and 12 dB compared to SC-FDMA and OFDMA,
respectively, at a BER of  $10^{-4}$.

\section{Message passing detection for OTFS-MA}
\label{sec4}
Although ML detection is optimal, its complexity grows exponentially 
with $M$ and $N$. In this section, we present a low complexity message 
passing based signal detection algorithm for OTFS-MA. Consider the 
OTFS-MA system model in \eqref{vecform}. Let $\Omega$ denote the support 
(positions of non-zeros) of the OTFS-MA transmit signal vector 
$[\mathbf{x}_1^T \mathbf{x}_2^T \cdots \mathbf{x}_{K_u}^T]^T$. Then, 
the system in \eqref{vecform} can be alternatively written as 
\begin{equation}
\mathbf{y} = \mathbf{H}\mathbf{x} + \mathbf{v},
\label{hx}
\end{equation}
where $\mathbf{H}=[\mathbf{H}_1\mathbf{H}_2\cdots \mathbf{H}_{K_u}]_{\Omega}$ 
is the channel restricted to $\Omega$ and 
$\mathbf{x}=[\mathbf{x}_1^T \mathbf{x}_2^T\cdots\mathbf{x}_{K_u}^T]^T_{\Omega}$
is the non-zero part of the OTFS-MA transmit vector. Then, \eqref{hx} can 
be modeled as a sparsely connected factor graph with $NM$ variable nodes 
corresponding to $\mathbf{x}$ and $NM$ observation nodes corresponding to 
$\mathbf{y}$. Denoting the support of the $s$th row of $\mathbf{H}$ by 
$\varphi_s$ and the support of the $r$th column of $\mathbf{H}$ by 
$\varphi_r$, each observation node $y_s$ is connected to the set of 
variable nodes \{$x_t, t \in \varphi_s$\}, and each variable node $x_r$ 
is connected to the set of observation nodes \{$y_t, t \in \varphi_r$\}. 
With this, the maximum a posteriori (MAP) detection rule for estimating 
the transmitted signal vector $\mathbf{x}$ is given by
\begin{equation}
\hat{\mathbf{x}}=\argmax_{\mathbf{x}\in \mathbb{A} ^ {NM}} \mbox{Pr}(\mathbf{x}|\mathbf{y},\mathbf{H}).
\label{mapx}
\end{equation}
The joint MAP detection in (\ref{mapx}) has exponential complexity. Hence, 
we use symbol by symbol MAP rule for $ 0 \leq r \leq NM-1 $ for detection as 
follows: 
\begin{eqnarray}
\label{MAPsbsmimo}
\hat{x}_r & = & \argmax_{a_j \in \mathbb{A}} \mbox{Pr}(x_r = a_j | \mathbf{y},\mathbf{H}) \nonumber \\
& = & \argmax_{a_j \in \mathbb{A}} {1 \over |\mathbb{A}|} \mbox{Pr}(\mathbf{y}|x_r =a_j , \mathbf{H}  ) \nonumber \\
& \approx & \argmax_{a_j \in \mathbb{A}} \prod_{t \in \varphi_r} \mbox{Pr}(y_t|x_r =a_j, \mathbf{H}).
\end{eqnarray}
Since the transmitted symbols can be assumed to be equally likely and the 
components of $\mathbf{y}$ can be assumed to be nearly independent for a 
given $x_r$, due to the sparsity in $\mathbf{H}$, (\ref{MAPsbsmimo}) can 
be solved using a message passing (MP) based approach. The message that is 
passed from the variable node $x_r$, for each $r= \{0,1,\cdots,NM-1\}$, to 
the observation node $y_s$ for $s\in \varphi_r$, is the pmf denoted by 
$\textbf{p}_{rs}=\{p_{rs}(a_j)|a_j \in \mathbb{A}\}$ of the symbols in 
the constellation $\mathbb{A}$. The steps involved in message passing 
detection can be described as follows:
\begin{algorithmic}[1]
\label{alg1}
\State \textbf{Inputs}: $\mathbf{y}$, $\mathbf{H}$, $n_{max} $: maximum 
number of iterations.
\State \textbf{Initialization}: Iteration index $k=0$, pmf 
$\mathbf{p}_{rs}^{(0)}=1/|\mathbb{A}| \ \forall \ r \in \{0,1,\cdots,NM-1 \}$ 
and $s \in \varphi_r $.
\State \textbf{Messages from $y_s$ to $x_r$}: The message passed from 
$y_s$ to $x_r$ is a Gaussian pdf which can be computed from
\begin{equation}
y_s= x_rH_{s,r}+\underbrace{\sum_{t \in \varphi_s,t \neq r } x_tH_{s,t} + v_s}_{\text I_{sr}} .
\end{equation}
The interference plus noise term $I_{rs}$ is approximated as a Gaussian 
r. v. with mean and variance given by
\begin{small}
\begin{equation*}
\mu_{sr}^{(k)} = \mathbb{E}[I_{sr}] = \sum_{t \in \varphi_s , t \neq r } \sum_{j=1}^{|\mathbb{A}|} p_{ts}^{(k)}(a_j) a_j H_{s,t},
\end{equation*}
\begin{align*}
&( \sigma_{sr}^{(k)})^2 = \text{Var}[I_{sr}] \nonumber \\ 
&= \sum_{\substack{t \in \varphi_s \\ t \neq r}} \Bigg( \sum_{j=1}^{\mathbb{|A|}}p_{ts}^{(k)}(a_j)|a_j|^2 |H_{s,t}|^2 - \bigg| \sum_{j=1}^{\mathbb{|A|}}p_{ts}^{(k)}(a_j)a_jH_{s,t} \bigg|^2 \Bigg)
\end{align*}
\hspace{4.5mm} $+ \ \sigma^2.$
\end{small}

\State \textbf{Messages from $x_r$ to $y_s$}: Message passed from variable 
nodes $x_r$ to observation nodes $y_s$ is the pmf vector 
$\textbf{p}_{rs}^{(k+1)}$ with the entries given by
\begin{equation}
p_{rs}^{(k+1)}=\Delta \ p_{rs}^{(k)}(a_j)+(1- \Delta) \  p_{rs}^{(k-1)}(a_j),
\end{equation}
where $\Delta \in (0,1]$ is the damping factor for improving convergence 
rate, and
\begin{equation}
p_{rs}^{(k)} \propto \prod_{t \in \varphi_r , t \neq s} \text{Pr}(y_t|x_r=a_j,\mathbf{H}),
\end{equation}
where 
\begin{equation*}
\text{Pr}(y_t|x_r=a_j,\mathbf{H}) \propto \text{exp} \Bigg( {-|y_t - \mu_{tr}^{(k)}-H_{t,r}a_j|^2 \over \sigma_{t,r}^{2(k)}}  \Bigg).
\end{equation*}

\State \textbf{Stopping criterion}: Repeat steps 3 and 4 till 
$\max\limits_{r,s,a_j} |p_{rs}^{(k+1)}(a_j) - p_{rs}^{(k)}(a_j)| < \epsilon$ 
(where $\epsilon$ is a small value) or the maximum number of iterations, 
$n_{max}$, is reached.
\State \textbf{Output}: Output the detected symbol as
\begin{equation}
\hat{x}_r=\argmax_{a_j \in \mathbb{A}}p_r(a_j), \:\: r\in {0,1,2,\cdots, NM-1},
\end{equation}
where
\begin{equation}
p_r(a_j) = \prod_{t \in \varphi_r} \text{Pr}(y_t|x_r =a _j,\mathbf{H}).
\end{equation}
\end{algorithmic}

\subsection{BER performance results with MP detection}
\label{sec4a}
\begin{figure}
\centering
\includegraphics[width=9.5cm, height=6.5cm]{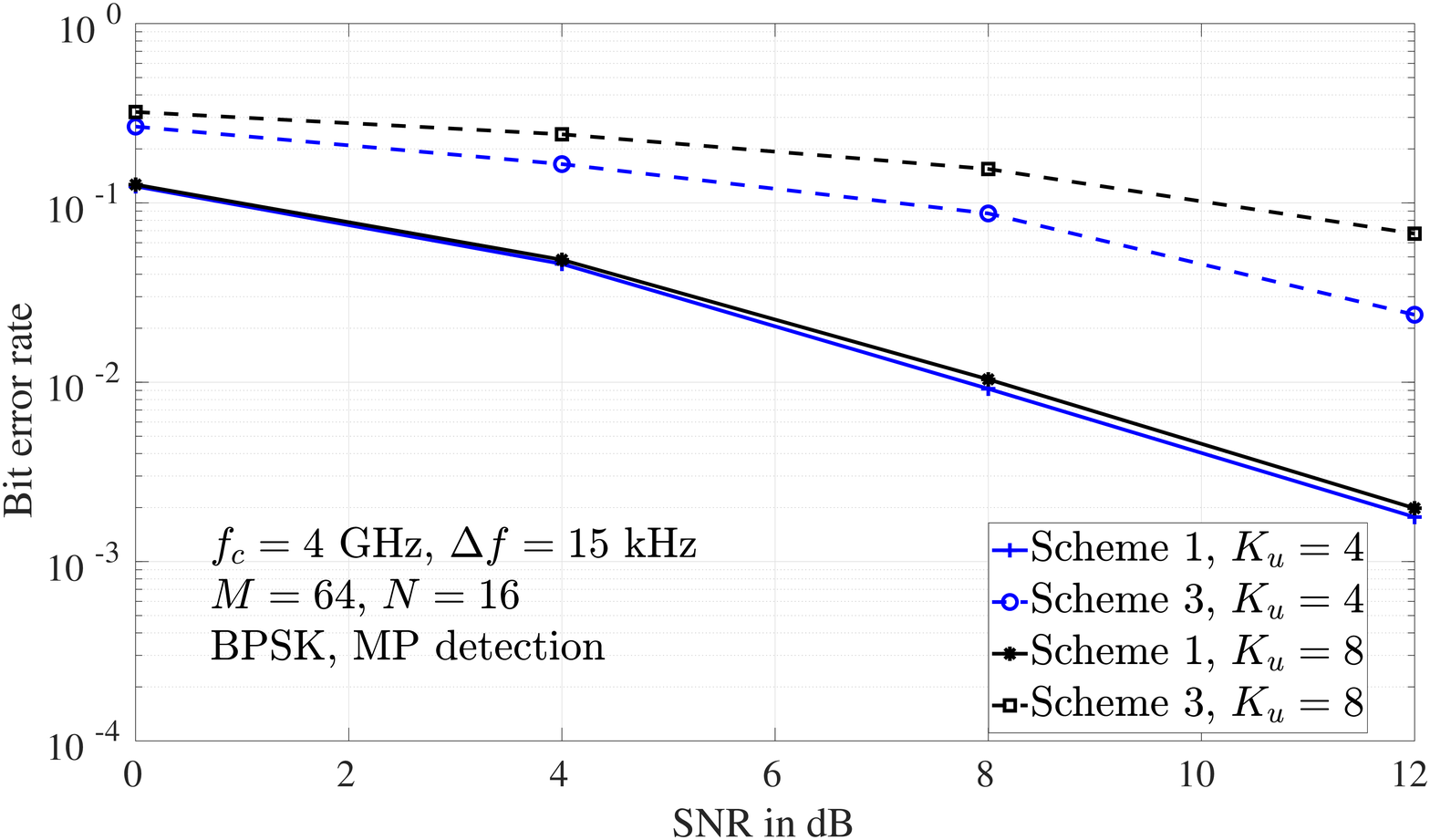}
\vspace{-3mm}
\caption{BER performance of uplink OTFS-MA with DDRB allocation Schemes 1 
and 3 with $K_u=4,8$ users, $M=64$, $N=16$, and MP detection.}
\vspace{-4mm}
\label{ber_mp1}
\end{figure}

Figure \ref{ber_mp1} shows the BER performance of OTFS-MA with DDRB 
allocation Schemes 1 and 3 using MP detection. All the systems 
considered use a carrier frequency of 4 GHz, a subcarrier spacing of 
15 kHz, and BPSK modulation. For all the users, we have considered
a 10-tap channel with exponential power delay profile and Jakes Doppler 
spectrum. The delay taps considered for each user's channel is
$\tau_{u,i}=[0,\; 1.04,\; 2.08,\; 3.12,\; 4.16,\; 5.2,\; 6.25,\; 7.29,\; 
8.33,\; 9.37]\; \mu$s $\forall u\in \lbrace 0,1,\cdots K_u-1 \rbrace$. 
The Doppler shift corresponding to the $i$th tap of $u$th user's channel is 
generated using $\nu_{u,i}=\nu_{\mbox{\scriptsize{max}}}\cos(\theta_{u,i})$,
where $\nu_{\mbox{\scriptsize{max}}}$ is the maximum Doppler shift and
$\theta_{u,i}$ is uniformly distributed over $[-\pi,\pi]$. The maximum 
Doppler shift considered is 1 kHz for all the users which corresponds to 
a velocity of 270 km/h. We have used a delay-Doppler grid with $M=64$ and 
$N=16$ and plotted the BER performance of Schemes 1 and 3 with $K_u=4$ and 
$8$, using the MP detection. From the Fig. \ref{ber_mp1}, we observe that 
the performance of Scheme 1 is superior compared to that of Scheme 3. 
Also, the performance of Scheme 1 does not degrade with the increase
in the number of uplink users, whereas the performance of Scheme 3
degrades with the increase in the number of users, as observed
with ML detection in Sec. \ref{sec3}.

\begin{figure}
\centering
\includegraphics[width=9.5cm, height=6.5cm]{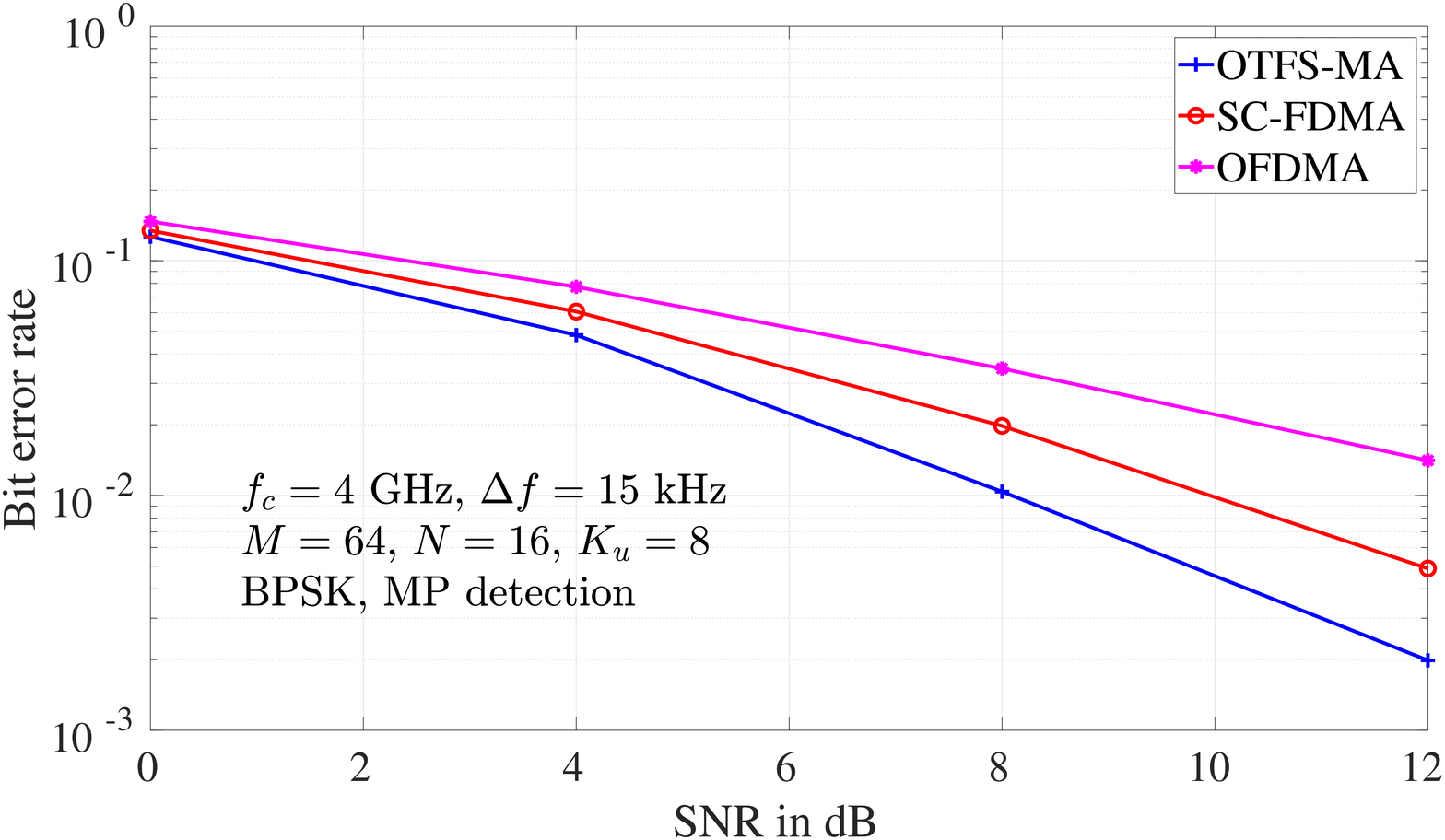}
\vspace{-3mm}
\caption{BER performance comparison between OTFS-MA, OFDMA, and 
SC-FDMA with MP detection.} 
\vspace{-4mm}
\label{ber_mp2}
\end{figure}

{\em Comparison between OTFS-MA, OFDMA, and SC-FDMA:}
Figure \ref{ber_mp2} shows the BER performance of OTFS-MA with allocation
Scheme 1, OFDMA, and SC-FDMA using message passing detection. OTFS-MA 
uses an $N\times M=16\times 64$ delay-Doppler grid which is allocated to 
$K_u=8$ users. All the systems use 4 GHz carrier frequency and a subcarrier 
spacing of 15 kHz. The channel corresponding to each user is assumed to 
have ten taps ($P_u=10$, $\forall u$) with an exponential 
power delay profile and Jakes Doppler spectrum. All the other simulation 
parameters considered are the same as those used in Fig. \ref{ber_mp1}. 
From Fig. \ref{ber_mp2}, it can be seen that OTFS-MA achieves superior 
performance compared to OFDMA and SC-FDMA, reiterating the results 
obtained with ML detection in Sec. \ref{sec3}. 

\begin{figure*}[t]
\centering
\includegraphics[width=16 cm, height=5.5 cm]{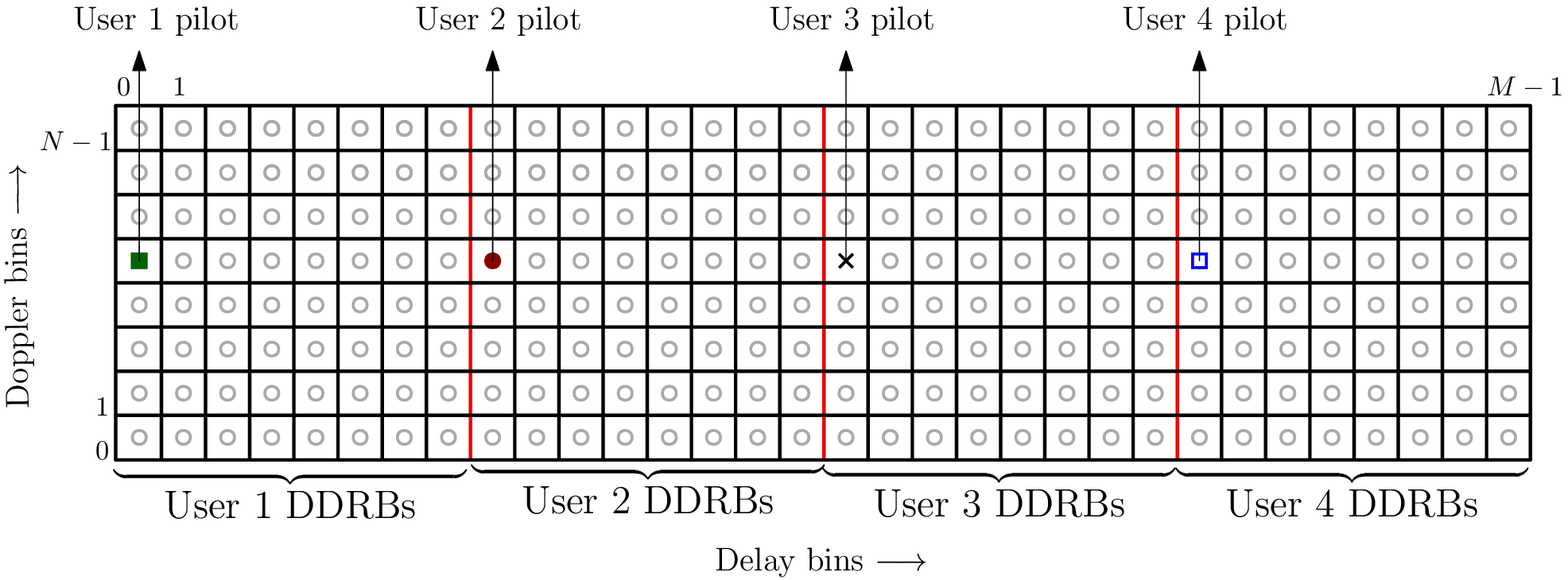}
\caption{Multiuser pilot placement on $N\times M$ delay-Doppler grid 
for Scheme 1 (`o' indicates zeros).}
\label{pilot_alloc_1}
\vspace{-0mm}
\end{figure*}

\begin{figure*}[t]
\centering
\includegraphics[angle=89.99, width=17 cm, height=5.5 cm]{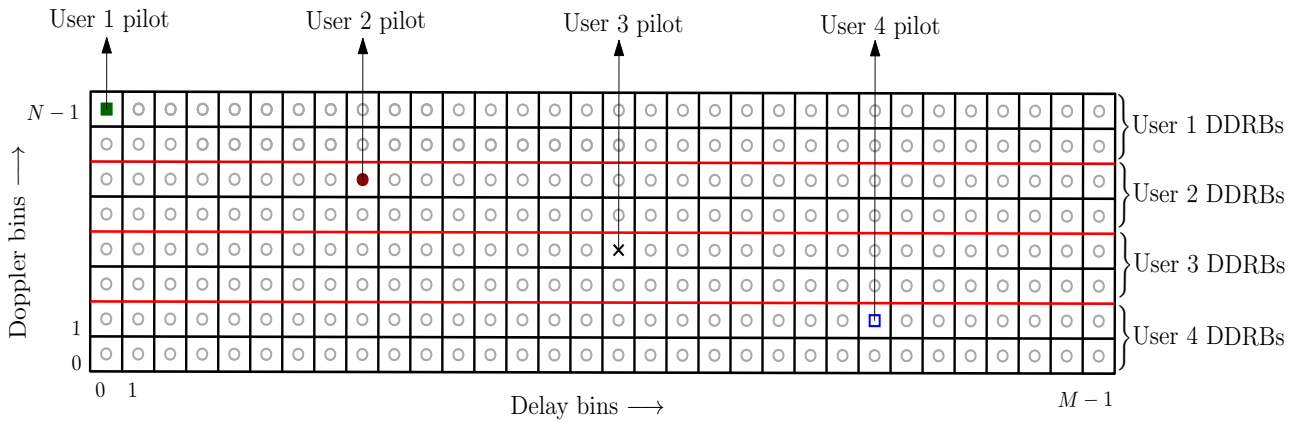}
\caption{Multiuser pilot placement on $N\times M$ delay-Doppler
grid for Scheme 2 (`o' indicates zeros).}
\label{pilot_alloc_2}
\vspace{-2mm}
\end{figure*}

\section{Channel estimation in OTFS-MA}
\label{sec5}
In this section, we present a channel estimation technique for uplink 
OTFS-MA with DDRB allocation Schemes 1 and 2. This technique uses an 
impulse function ($\delta(k,l)$) in the delay-Doppler domain as the pilot. 
The pilot corresponding to each user is placed in the delay-Doppler grid 
such that they can be received without interference at the BS. The pilot 
corresponding to the $u$th user is an impulse denoted by 
$\delta (k_u^p,l_u^p)$, such that the point $(k_u^p,l_u^p)$ is a DDRB 
allocated to the $u$th user. Each user's pilot has a space reserved around 
it  in the delay-Doppler plane to account for the maximum delay and Doppler
spread of the channel. Since the transmitted pilots are impulse functions, 
they are spread by the channel to the extent of 
the support of each user's channel in the delay-Doppler domain. Hence, if 
the pilots are placed sufficiently far apart in the delay-Doppler plane, 
they can be received at the BS without interference. For the placement of 
pilots of different users, we take into account fractional Dopplers in 
the channels as in \eqref{inpoutfracdop}. From \eqref{inpoutfracdop}, it 
can be seen that due to the fractional Doppler values, the channel spreads 
completely along the Doppler domain \cite{emb_pil}. Hence, the pilot 
corresponding to the $u$th user, denoted by $x^p_u[k,l]$ is placed such that
\begin{equation}
x^p_u[k,l]=\begin{cases}
1  & \textnormal{if} \; k=k^p_u, l=l^p_u  \\
0  & \textnormal{otherwise} .
\end{cases},
\end{equation}
where $(k^p_u,l^p_u)$ is a DDRB allocated to the $u$th user and 
$l^p_{u+1}-l^p_u>\max \limits_i 
(\alpha_{u,i})\; \textnormal{for every} \; u$. Note that this requires
$M/K_u>\max \limits_{u,i}(\alpha_{u,i})$ for Scheme 1 and $M>\sum
\limits_{u}\max \limits_{i}(\alpha_{u,i})$ for Scheme 2. The interaction of 
pilot with the channel results in a 2D convolution of the delay-Doppler 
impulse response with the pilot. The received pilot corresponding to the $u
$th user can be written using \eqref{2Dconv} as
\begin{align}
y^p_u[k',l']=&\frac{1}{MN}\tilde{h}_u[(k'-k^p_u)_N,(l'-l^p_u)_M]+v[k',l'],
\label{chan_est}
\end{align}
which gives the estimated channel gains of the $u$th user, where
$k'\in \{0,1,\cdots,N-1\}$ and
$l'\in \{l_u^p,\cdots,l_u^p+\max\limits_i \alpha_{u,i}\}$. 
Figures \ref{pilot_alloc_1} and \ref{pilot_alloc_2} illustrate one of 
the possible ways of placing the pilots on the $N\times M=8\times 32$ 
delay-Doppler grid allocated to $K_u=4$ users for Schemes 1 and 2, 
respectively, taking $\max\limits_{u,i} \alpha_{u,i}=7$.

\subsection{Performance results}
\begin{figure}
\centering
\includegraphics[width=9.5cm, height=6.5cm]{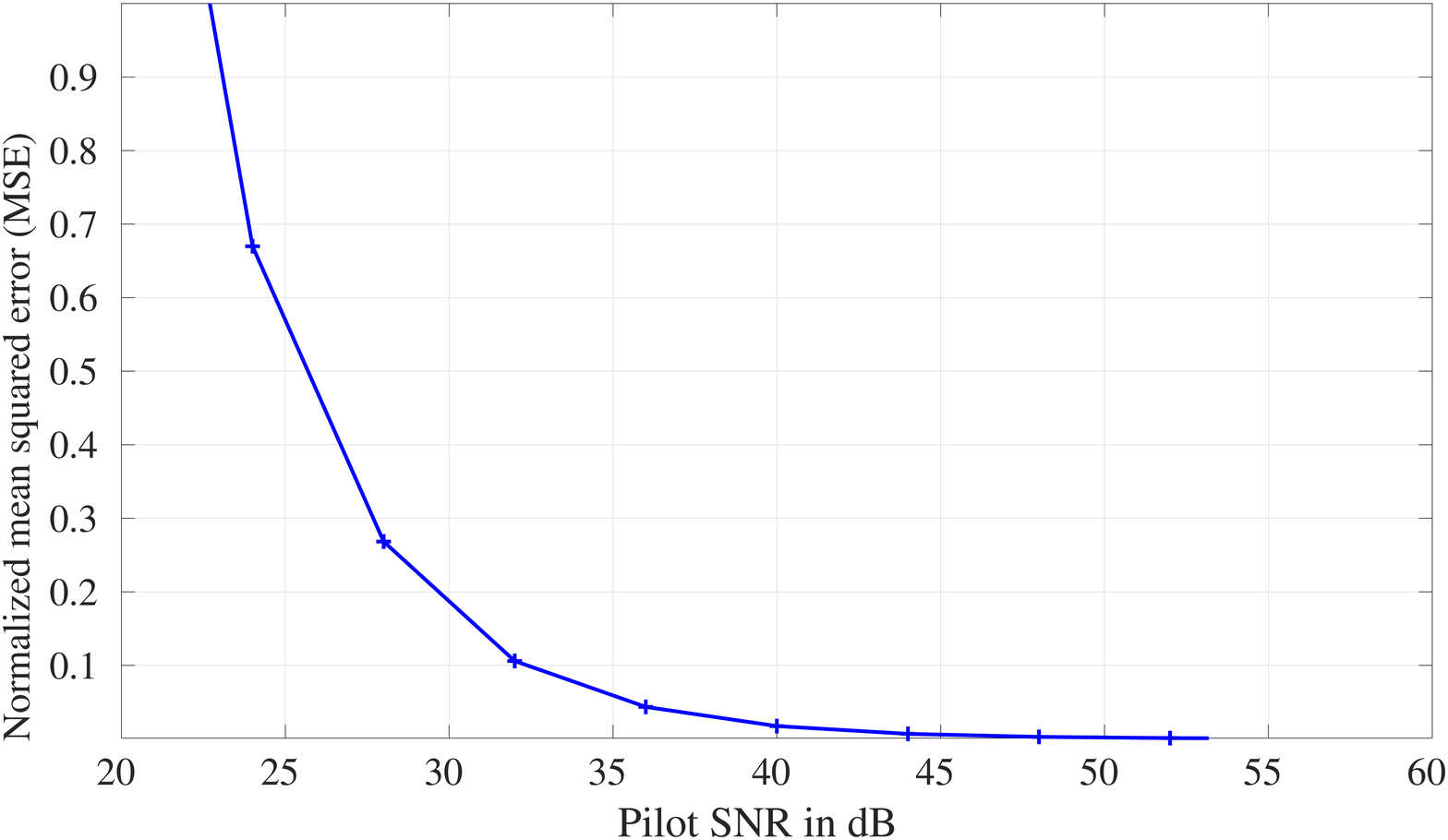}
\vspace{-3mm}
\caption{Normalized mean squared error of the estimated channel in 
uplink OTFS-MA.} 
\vspace{-4mm}
\label{mse}
\end{figure}

\begin{figure}
\centering
\includegraphics[width=9.5cm, height=6.5cm]{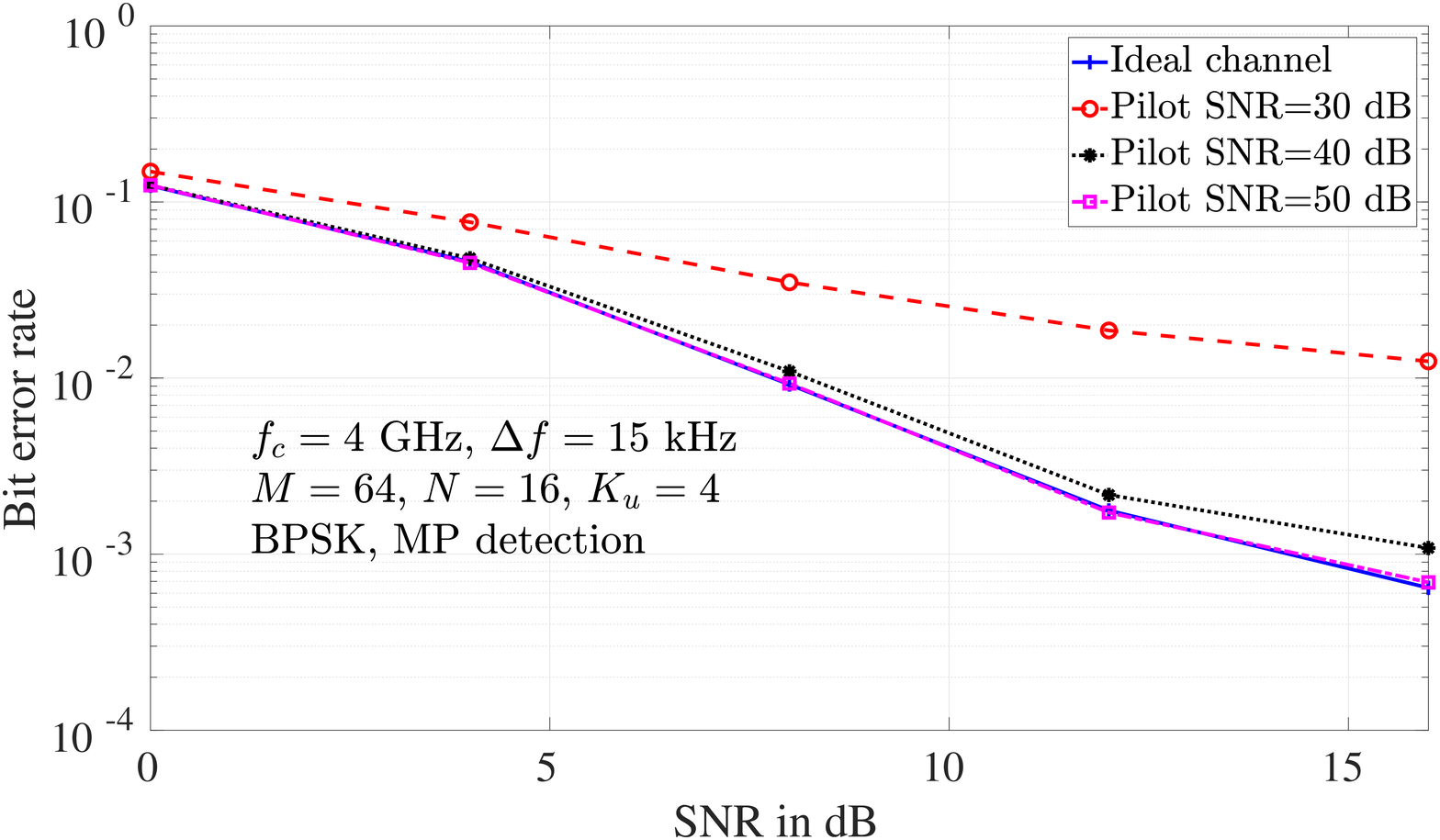}
\vspace{-3mm}
\caption{BER performance of OTFS-MA with estimated channel in uplink 
OTFS-MA.} 
\vspace{-4mm}
\label{ber_ce}
\end{figure}

Figure \ref{mse} shows the normalized mean squared error (MSE) of the
estimated channel as a function of pilot SNR for four users ($K_u=4$) 
in uplink OTFS-MA with Scheme 1 of DDRB allocation. The channel 
corresponding to each user is assumed to have ten taps $(P_u=10, \forall u)$ 
with an exponential power delay profile and Jakes Doppler spectrum. All 
the other channel parameters are same as considered for Fig. \ref{ber_mp1}. 
The pilots are placed on $N\times M=16 \times 64$ delay-Doppler grid for 
the channel estimation. From Fog, \ref{mse}, we see that the normalized 
MSE decreases with the increase in pilot SNR and the MSE is less than 
0.01 for pilot SNR larger than 36 dB. In Fig. \ref{ber_ce}, we plot the
BER performance of OTFS-MA with Scheme 1 allocation with $K_u=4$ using 
the channel estimation scheme described above and MP detection for 
different values of pilot SNR. The channel is estimated during the
pilot frame which is used for detection in the subsequent data frame. 
From Fig. \ref{ber_ce}, it can be observed that the BER performance 
achieved with the estimated channel is close to the performance with 
perfect channel knowledge for pilot SNRs of 40 and 50 dB. 

\section{Conclusions}
\label{sec6}
We considered the problem of multiple access using the recently 
proposed OTFS modulation for multiuser communication on the uplink. 
Three different schemes to allocate delay-Doppler resource blocks to 
the users were considered. The BER performance of OTFS-MA in comparison
with those of OFDMA and SC-FDMA was investigated considering ML detection 
for small dimension systems and message passing detection for large 
dimension systems. OTFS-MA was found to achieve better performance 
compared to OFDMA and SC-FDMA on the uplink in high mobility environments. 
Also, a pilot based channel estimation scheme in the delay-Doppler domain 
for OTFS-MA was shown to achieve a performance close to that with perfect 
channel knowledge.

\end{document}